\documentclass{nature}
\usepackage[english]{babel}
\usepackage[T1]{fontenc}
\usepackage{amsmath}
\usepackage{amssymb}
\usepackage{bm}
\usepackage{babel}
\usepackage{wasysym}
\usepackage{graphicx}
\usepackage{subfigure}
\usepackage{multirow}
\usepackage{color}
\usepackage{ulem}

\usepackage[font=small,labelfont=bf,labelsep=space]{caption}
\definecolor{blue}{rgb}{0,0,1}

\definecolor{darkgreen}{RGB}{0,128,0}

\begin{document}
\title{Robustness of momentum-indirect interlayer excitons in MoS$_2$/WSe$_2$ heterostructure against charge carrier doping}
\author{Ekaterina Khestanova$^{1*}$, Tatyana Ivanova$^2$, Roland Gillen$^3$, Alessandro D'Elia$^4$, Oliver Nicholas Gallego Lacey$^4$, Lena Wysocki$^4$, Alexander~Gr\"uneis$^{4, 5}$, Vasily Kravtsov$^2$, Wlodek Strupinski$^6$, Janina Maultzsch$^3$, Viktor Kandyba$^7$, Mattia Cattelan$^{7, 8}$, Alexei Barinov$^7$, Jos\'e Avila$^9$, Pavel Dudin$^9$, Boris V.~Senkovskiy$^{4*}$}
\maketitle
\begin{affiliations}
\item ICFO-Institut de Ciencies Fotoniques, The Barcelona Institute of Science and Technology, Castelldefels (Barcelona) 08860, Spain
\item School of Physics and Engineering, ITMO University, 197101 Saint Petersburg, Russia
\item Department Physik, Friedrich-Alexander-Universität Erlangen-N{\"u}rnberg, Staudtstra{\ss}e 7, 91058 Erlangen, Germany
\item II. Physikalisches Institut, Universit\"at zu K\"oln, Z\"ulpicher Stra{\ss}e 77, 50937 K\"oln, Germany
\item Institut für Festkörperelektronik, Technische Universität Wien,
Gußhausstraße 25-25a, 1040 Wien, Austria
\item Faculty of Physics, Warsaw University of Technology, 00-662 Warsaw, Poland
\item Elettra-Sincrotrone Trieste S.C.p.A., Basovizza, 34149 Trieste, Italy
\item Department of Chemical Sciences, University of Padova, Via F. Marzolo 1, 35131 Padova, Italy
\item Synchrotron SOLEIL, Universit\'e Paris-Saclay, L’Orme des Merisiers Saint-Aubin, Gif sur Yvette, France

\end{affiliations}

\pagebreak

\section*{Abstract}
 Monolayer transition-metal dichalcogenide (TMD) semiconductors exhibit strong excitonic effects and hold promise for optical and optoelectronic applications. Yet, electron doping of TMDs leads to the conversion of neutral excitons into negative trions, which recombine predominantly non-radiatively at room temperature. As a result, the photoluminescence (PL) intensity is quenched. Here we study the optical and electronic properties of a MoS$_2$/WSe$_2$ heterostructure as a function of chemical doping by Cs atoms performed under ultra-high vacuum conditions. By PL measurements we identify two interlayer excitons and assign them to the momentum-indirect Q-$\Gamma$ and K-$\Gamma$ transitions. The energies of these excitons are in a very good agreement with \textit{ab initio} calculations. We find that the Q-$\Gamma$ interlayer exciton is robust to the electron doping and is present at room temperature even at a high charge carrier concentration ($\sim$ 10$^{13}$ cm$^{-2}$). Submicrometer angle-resolved photoemission spectroscopy ($\mu$-ARPES) reveals charge transfer from deposited Cs adatoms to both the upper MoS$_2$ and the lower WSe$_2$ monolayer without changing the band alignment. This leads to a small ($\sim$ 10 meV) energy shift of interlayer excitons. Robustness of the momentum-indirect interlayer exciton to charge doping opens up an opportunity of using TMD heterostructures in {light-emitting devices 
 that can work at room temperature at high densities of charge carriers}.

\maketitle

\section*{Introduction}
Monolayer transition metal dichalcogenides (TMDs) are direct band gap semiconductors with multi-valley electronic band structure and strong light-matter coupling. These properties give rise to a variety of electron-hole pair excitations (excitons) that are stable at room temperature.~\cite{Mak2016} When monolayer TMDs are stacked in van der Waals heterostructures with type-II band alignment, electrons and holes reside in different layers and interlayer excitons (ILEs) emerge. Such ILEs demonstrate extremely long lifetime, large valley polarization and coherence, necessary for valleytronics applications.~\cite{Rivera2015, Rivera2016, Miller2017} {It was demonstrated that application of external electric fields provides control over properties of ILEs: their intensity and photon energy,~\cite{Rivera2015} valley polarisation,~\cite{Rivera2016, Ciarrocchi2019} and flux.~\cite{Unuchek2018} Such versatile tunability of ILEs makes TMDs heterostructures promising} for designing optoelectronic devices such as light-emitting diodes (LEDs), solar cells, photodetectors, energy conversion and storage devices. First, an ILE in MoS$_2$/WSe$_2$ heterostructures fabricated on SiO$_2$ substrates was observed with light emission at around 1.6 eV,~\cite{Fang2014,Kunstmann_Nature18, Unuchek2018} and the electrical control of its diffusion in a sample encapsulated between hexagonal boron nitride (hBN) layers was also demonstrated.~\cite{Unuchek2018} Experimental data and calculations suggest that this ILE is momentum-indirect, i.e. an electron in MoS$_2$ and a hole in WSe$_2$ have different wavevectors.~\cite{Kunstmann_Nature18} 
It was proposed that this ILE consists of an electron residing in the K valley of MoS$_2$ and a hole residing in the $\Gamma$ valley of WSe$_2$.~\cite{Kunstmann_Nature18}  Further, a momentum-direct ILE, coupling the K valleys of MoS$_2$ and WSe$_2$, was found in  MoS$_2$/WSe$_2$ heterostructure at around 1 eV.~\cite{Karni_PRL2019} Interestingly, in a very similar system -- hBN-encapsulated MoS$_2$/WS$_2$, two momentum-indirect ILEs were observed.~\cite{Okada2018} These ILEs are associated with two conduction band (CB) valleys of MoS$_2$ located in the K and Q points and the valence band (VB) valley of WS$_2$ at the $\Gamma$ point.~\cite{Okada2018} The hBN encapsulation plays an important role in this finding. For example, for a non-encapsulated MoS$_2$/WS$_2$ heterostructure on SiO$_2$, the contribution of two momentum-indirect ILEs cannot be well resolved.~\cite{Wurstbauer_PRB20} These results suggest further investigation of the MoS$_2$/WSe$_2$ system, in which two types of momentum-indirect ILEs can be also present.

Besides the origin of ILEs in TMD heterostructures, another open question is the impact of charge doping on the ILE emission. This is an important question not only from a fundamental but also from an applied point of view. For devices that require both light emission and charge transport, it is important to maintain the radiative efficiency at an increased charge carrier densities. For example, conductivity and the charge carrier mobility in TMDs can be increased by %\sout{electron}{\blue }
charge doping.~\cite{Baugher2013} However, increased charge carrier concentration in TMD monolayers leads to the conversion of neutral excitons into %\sout{the negative trions (composed of two electrons and one hole)} {\blue}
trions, which have long radiative and short non-radiative lifetimes.~\cite{Lien468} %{\blue}
In the case of electron doping, negative trions, composed of two electrons and one hole, will appear. Here, the main mechanism responsible for non-radiative recombination is the Auger process, where the energy of  electron–hole recombination is transferred to the third particle (i.e. an electron in negative trions).~\cite{Kurzmann_Auger_NL, Lien468, Carmiggelt_SciRep2020} Thus, Auger recombination dramatically decreases the PL quantum yield, which complicates the use of TMDs as optical active elements at a high concentration of charge carriers. Nevertheless, the impact of electron doping on PL of momentum-indirect ILEs involving different valleys in CB might be different.
To answer these questions, we explore a high-quality MoS$_2$/WSe$_2$/hBN van der Waals heterostructure and find two momentum-indirect ILEs. To understand the origin of these ILEs, we perform \textit{ab initio} calculations of optical transitions in MoS$_2$/WSe$_2$ system. 
We investigate the effect of electron doping on the electronic and optical properties of the MoS$_2$/WSe$_2$ heterostructure using the deposition of Cs atoms \textit{in situ}. Ultra-high-vacuum (UHV) PL spectroscopy reveals that the Q-$\Gamma$ exciton is robust to the electron doping and is emitting light even at high charge carrier concentration ($\sim$ 10$^{13}$ cm$^{-2}$) at room temperature. {Using the spatially-resolved angle-resolved photoemission spectroscopy with sub-micron spot ($\mu$-ARPES), we} find that the deposition of Cs has a minor impact on the VB alignment in MoS$_2$/WSe$_2$ heterostructure. This is due to the intercalation of Cs in the van der Waals gap, which leads to equal charge distribution between the TMD monolayers. This explains the minor shift of ILEs in the MoS$_2$/WSe$_2$ system upon Cs doping. The robustness of momentum-indirect ILEs to the charge doping paves the way towards the design of novel optoelectronic devices based on TMD heterostructures with combined effective light emission and improved charge transport properties at high charge carrier concentrations.

\section*{Sample fabrication and characterization}
PL and $\mu$-ARPES studies were performed on similar samples fabricated by the deterministic transfer technique.~\cite{Castellanos-Gomez2014} The MoS$_2$/WSe$_2$ heterobilayers were assembled layer by layer on hBN flakes placed on epitaxial bilayer graphene grown on SiC. The advantage of such a substrate is that hBN is an atomically flat and pure insulator (in contrast to SiO$_2$), while graphene provides the electrical conductivity (through the TMD edges lying outside the hBN flake), necessary for $\mu$-ARPES measurements. The hBN flake also ensures the absence of charge transfer between the TMD heterostructure and graphene in PL measurements. To minimize organic residues, we exfoliated monolayer TMDs on ultraviolet-ozone cleaned polydimethylsiloxane (PDMS) stamps.~\cite{Novotny_PDMScleaning} Post-transfer thermal annealing of the TMDs heterostructure is an important step to provide strong interlayer coupling.~\cite{Tongay2014}
Our samples were annealed in UHV at 250 $^{\circ}$C for 3 hours.For further details see Supplementary Information S1.

\begin{spacing}{1.0}
\begin{figure} \centering \includegraphics[height=9.5 cm,
    keepaspectratio]{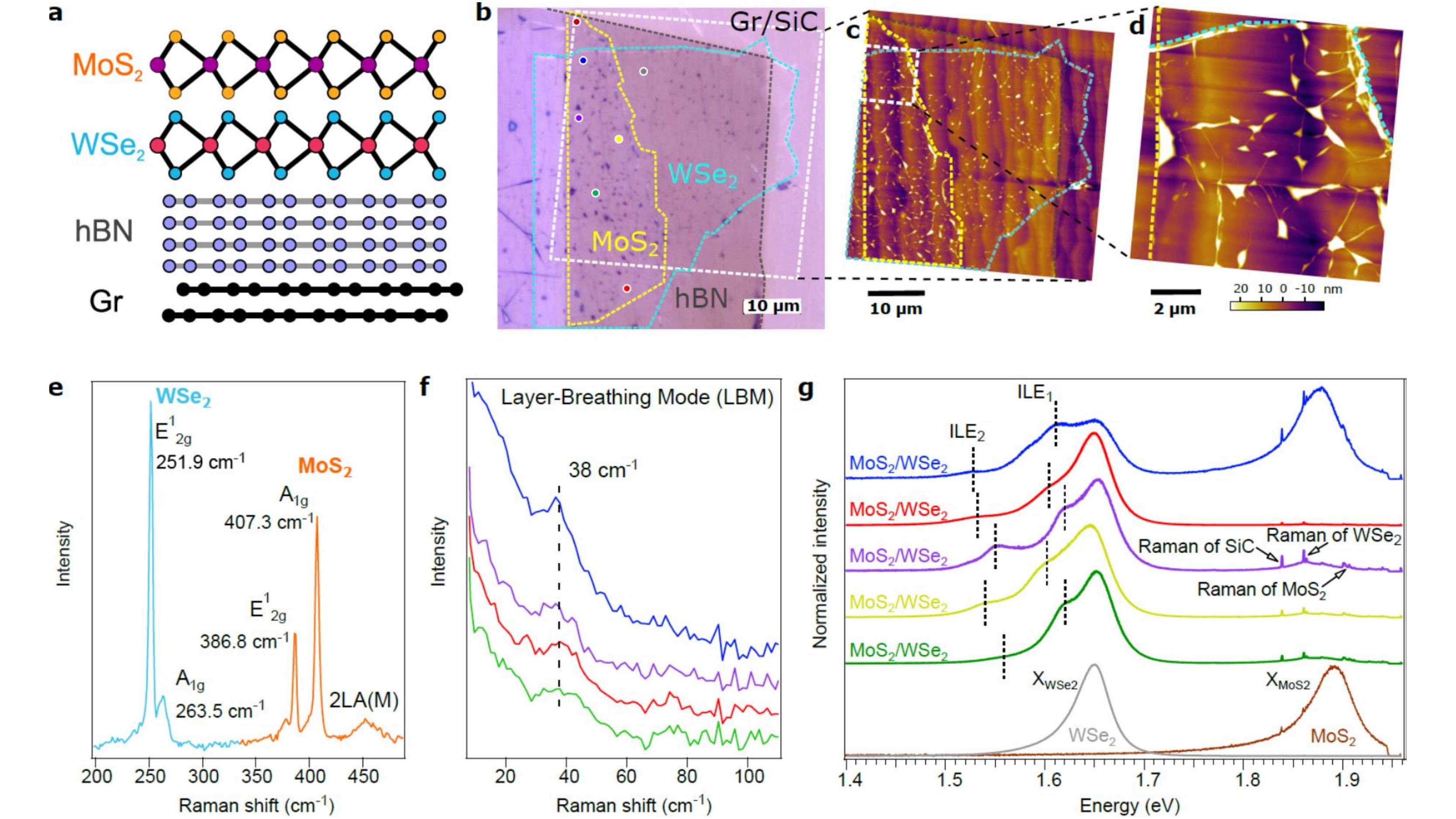} \caption{(a) Illustration of the MoS$_2$/WSe$_2$ heterostructure on hBN on epitaxial bilayer graphene. (b) Optical microscope image of the aligned MoS$_2$/WSe$_2$ heterostructure. The edges of TMD monolayers are indicated. The white dashed line shows the area where the AFM image was measured. {The coloured dots indicate the positions of the laser during measurements shown in Fig.1 (g).} (c) AFM topography image of the region outlined by the white dashed line in the panel b. (d) Zoomed-in AFM topography image of the region outlined by the white dashed line in the panel c. (e) Raman spectrum of the heterostructure measured with a 532 nm laser. Vibrational modes of MoS$_2$ and WSe$_2$ are indicated. (f) Low-frequency Raman spectra of the heterostructure measured at different points with a 633 nm laser. LBM mode at 38 cm$^{-1}$ is indicated. (g) PL spectra of the heterostructure measured at room temperature at different points in the bubble-free regions using a 633 nm laser. In addition to the PL peaks from intralayer excitons of MoS$_2$ and WSe$_2$, two low-energy features are observed, designated as ILE$_1$ and ILE$_2$. The spectra of the individual monolayers are shown at the bottom for reference. 
    } \label{fig:Fig1} 
    \end{figure}
\end{spacing}     

Fig. 1 (a) illustrates the MoS$_2$/WSe$_2$ heterostructure on hBN on bilayer graphene/SiC. The optical image of the sample with indicated edges of the MoS$_2$ and WSe$_2$ monolayers is shown in Fig. 1 (b). The monolayers were not perfectly aligned. The twist angle between TMD monolayers was varying within 5$^{\circ}$ as determined using the second-harmonic generation (SHG) measurements (see Supplementary Information, Fig. S2).
%~\cite{Li2013, Kumar_PRB2013, Malard_PRB2013, Hsu2014} 
Our heterostructures are quite large -- several tens of microns, which allows us to probe and compare different points on the sample. The atomic force microscopy (AFM) measurements shown in Fig. 1 (c, d) reveal the surface topography. The exfoliated flakes follow the characteristic atomically flat terraces of hexagonal SiC(0001) with epitaxial bilayer graphene. UHV annealing leads to the aggregation of contaminants trapped between the layers into the bubbles (or blisters). These bubbles are also visible in optical microscope and we select the bubble-free regions to measure Raman and PL spectra with the laser spot of 1 $\mu$m. AFM reveals the height of MoS$_2$/WSe$_2$ heterobilayer of $\sim$ 1.5 nm (Supplementary information, Fig. S3), indicating the absence of organic residues at the interface and the surface.~\cite{Novotny_PDMScleaning} 

Fig. 1 (e) shows the typical Raman modes of MoS$_2$ and WSe$_2$ in the heterostructure region (532 nm laser). The clean interface provides strong interaction between TMD monolayers. Indeed, in Raman spectra measured with a low wavenumber filter (633 nm laser) we observe layer-breathing mode (LBM), Fig. 1 (f). This mode arises from the out-of-plane vibration of two monolayers when they are in close proximity. The frequency of the LBM mode (38 cm$^{-1}$) is comparable to that measured on a heterostructure fabricated from chemically grown TMDs on SiO$_2$ (35 cm$^{-1}$).~\cite{LBM_PRB2015} The small difference in frequency might be related to the different substrates or slightly different twist angles. 
 {When the laser spot is directed into the bubble, the LBM mode is absent due to weak interlayer coupling in these areas, the signal from the individual monolayers dominates the PL spectra and the ILE peaks, discussed below, are vanishing. These facts allow us to exclude the contribution of bubbles to the PL spectra.}
Fig. 1 (g) shows the PL spectra measured at room temperature in air in different bubble-free regions of the sample. One can identify the PL peak of MoS$_2$ at 1.88 eV. This peak is composed of intralayer neutral exciton and negatively charged trion (which manifests itself in the asymmetry of the PL peak towards lower energies).~\cite{Mak2013, Christopher2017} The PL of MoS$_2$ is suppressed in most of the assessed points due to the interlayer charge transfer in a type-II TMD heterostructure.~\cite{Hong2014, Fang2014, Rivera2015} The PL peak at 1.65 eV is the intralayer exciton of WSe$_2$, later referred to as X$_{\rm{WSe2}}$. The energy of the momentum-direct K-K exciton in MoS$_2$/WSe$_2$ system is in the infrared range.~\cite{Karni_PRL2019} Other multi-particle exciton complexes, which may appear energetically below the X$_{\rm{WSe2}}$, such as trions, biexcitons, defect bound excitons, moir{\'e} excitons can be observed only at cryogenic temperatures.~\cite{ Urbaszek_PRB2014, You2015, Clark2016, Li2018, Paur2019, Seyler2019, Tran2019} Consequently, the PL peaks that are observed only in the heterostructure region at room temperature and whose energy is just below the X$_{\rm{WSe2}}$ peak should be attributed to the momentum-indirect ILEs. In contrast to the previous studies, revealing a single ILE peak at $\sim$ 1.6 eV in the MoS$_2$/WSe$_2$ heterostructures on SiO$_2$,~\cite{Kunstmann_Nature18, Unuchek2018, Nagler_pssB_2019} we clearly observe two peaks -- at $\sim$ 1.61 eV and at $\sim$ 1.55 eV, and denote them as ILE$_1$ and ILE$_2$, respectively. The energies and intensities of the two ILEs peaks slightly vary from point to point and from sample to sample (see Supplementary Information, Fig. S1). This can be attributed to the variations of interlayer twist angle, interlayer spacing and doping.

\section*{\textit{In situ} chemical doping and UHV PL studies}
To avoid the effect of air adsorbates (e.g. water molecules, oxygen, hydrocarbons) and to investigate the impact of chemical doping on intrinsic optical properties of MoS$_2$/WSe$_2$ system, we performed PL studies in UHV conditions. For this we used a unique UHV optical setup for PL and Raman studies.~\cite{Senkovskiy_NL2017, Hell2018, Senkovskiy2018a, Ehlen_2018} Fig. 2 (a) shows the evolution of the UHV PL spectra of the sample upon deposition of Cs. 
The Cs dose was calibrated using a quartz crystal microbalance. The measurements show that intralayer excitons (X$_{\rm{MoS2}}$ and X$_{\rm{WSe2}}$) and the interlayer exciton ILE$_2$ are suppressed as the Cs dose increases. 
The PL of excitons is mainly suppressed due to their conversion to negative trions~\cite{Mak2013, Mouri2013} via electron donation by Cs atoms with the subsequent non-radiative recombination of trions through Auger processes.~\cite{Kurzmann_Auger_NL, Lien468, Carmiggelt_SciRep2020} 
Moreover, doping increases the dielectric screening of electron-hole Coulomb interaction, and hence reduces the binding energy and the oscillator strength of excitons,~\cite{Chernikov_PRL2015} which in turn lowers the radiative decay rate. In addition, in case of intercalation of Cs into the van der Waals gap, the interlayer distance will be increased and, consequently, the PL intensity will be reduced.
After deposition of $\sim$ 1 {\AA} of Cs one can see almost no signal from X$_{\rm{MoS2}}$, X$_{\rm{WSe2}}$ and ILE$_2$ peaks (purple spectrum in Fig. 2 (a)). At the same time, the %\sout{ILE$_1$ peak is relatively intense}{\blue Ç
intensity of the ILE$_1$ peak is relatively large. Even after increasing the Cs dose to $\sim$ 2 {\AA}, the ILE$_1$ peak remains clearly visible, while its intensity is reduced by a factor of $\sim$ 3 (light blue spectrum in Fig. 2(a)). 
{{blue} The Cs dosage above 2 {\AA} has much less effect on PL spectra, and we did not observe complete suppression of the ILE$_1$ peak.}
Fig. 2 (b) shows the fits of the PL spectra in the pristine and Cs-doped heterostructure. The energies of PL peaks are identified. The full width at half-maximum (FWHM) of these peaks are 19, 21 and 40 meV for X$\rm{_{WSe2}}$, ILE$_1$ and ILE$_2$, correspondingly. Such narrow emission lines at room temperature indicate the high quality of our samples.~\cite{Urbaszek_PRX} This is due to the atomically flat hBN substrate without charge impurities and the absence of unintended functionalization by air adsorbates in UHV conditions. 
{Further evidence of high-quality samples comes from the well-resolved electronic bands in $\mu$-ARPES spectra shown below, since ARPES is a surface-sensitive technique requiring atomically-clean surfaces and interfaces.}
Interestingly, upon doping, the ILE$_1$ peak becomes even narrower (lower spectrum in Fig. 2 (b)). The FWHM of ILE$_1$ is 13~meV at the maximum doping level. This suggests that there is no broadening of ILEs due to the inhomogeneous doping or exciton scattering.

\begin{spacing}{1.0}
%\singlespacing
\begin{figure} \centering \includegraphics[width=1\textwidth,
    keepaspectratio]{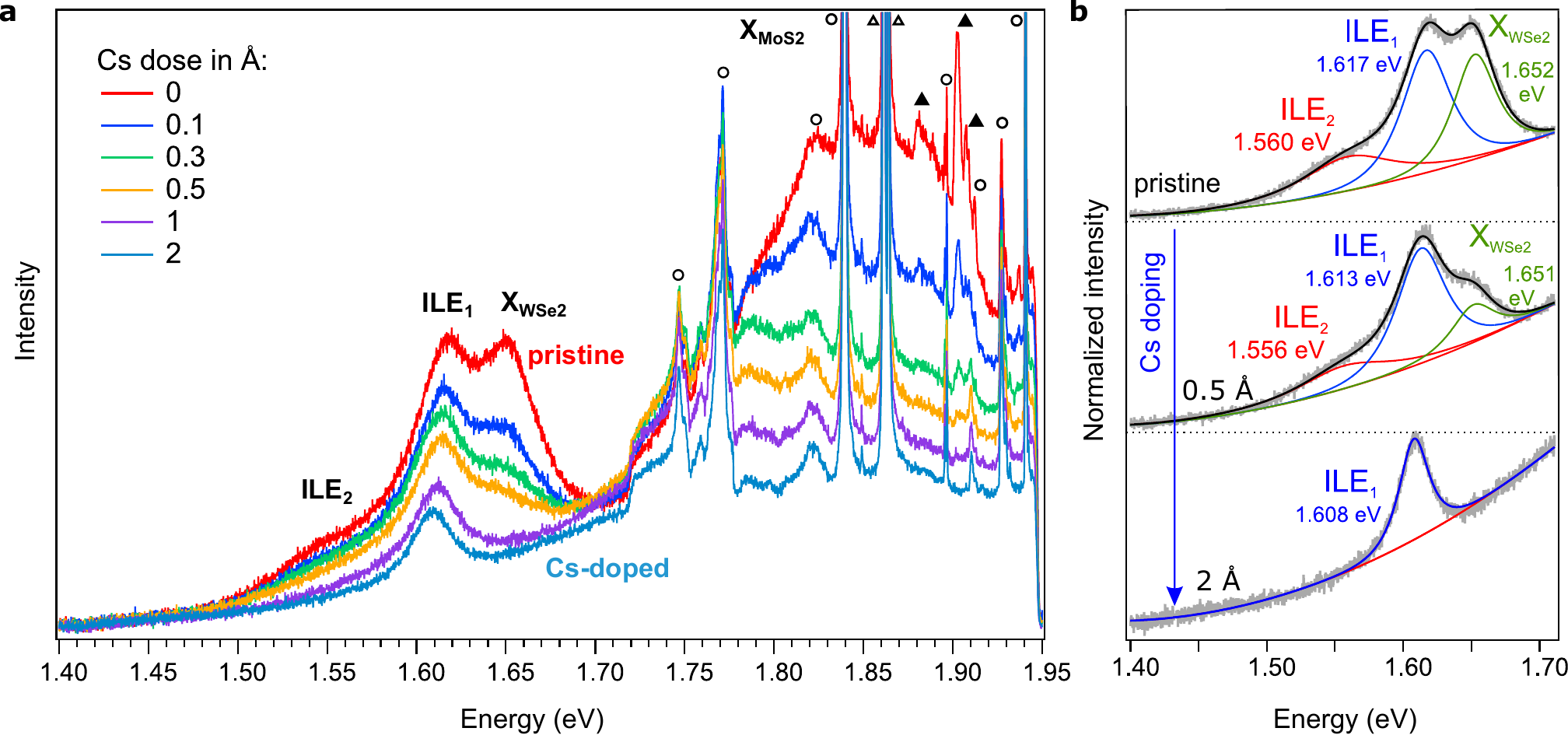} \caption{\label{fig:figure2} (a) UHV PL spectra of MoS$_2$/WSe$_2$ heterostructure on hBN on bilayer graphene/SiC upon Cs doping, room temperature, 633 nm laser. The intralayer and interlayer excitons are indicated as X$_{\rm{WSe2}}$ and ILE$_1$, ILE$_2$, correspondingly. The PL intensity of MoS$_2$ and WSe$_2$ (labeled as X$_{\rm{MoS2}}$ and X$_{\rm{WSe2}}$), as well as of intralayer ILE$_1$ exciton is completely suppressed as the Cs dose is increased to $\sim$ 2 {\AA}. At the same time, the intensity of the ILE$_1$ feature is suppressed by $\sim$ 3 times compared to the pristine sample. Raman peaks from SiC, MoS$_2$ and  WSe$_2$ are indicated by open circles, filled triangles and open triangles, respectively. 
    (b) Fits of the PL spectra of the pristine and doped heterostructure by mixed Gaussian-Lorentzian (0.2/0.8) function. The energies of the peaks are indicated.} \end{figure}
\end{spacing}

\section*{Energies of interlayer excitons}
To understand the nature of the observed ILEs, we compare the energy position of PL peaks with calculations. The energies of the X$_{\rm{MoS2}}$, X$_{\rm{WSe2}}$ and ILE$_2$ peaks (1.9 eV and 1.65 eV, and 1.55 eV, correspondingly) agree very well with the reported \textit{ab initio} calculations of intralayer MoS$_2$ and WSe$_2$ K-K transitions and interlayer K-$\Gamma$ transition, respectively.~\cite{Karni_PRL2019} Previously,~\cite{Kunstmann_Nature18} for the heterostructures on SiO$_2$ substrate, the ILE$_2$ peak was not observed, and the K-$\Gamma$ transition was assigned to the peak at $\sim$ 1.6 eV, which we designated as ILE$_1$. 

\begin{figure} 
\centering 
\includegraphics[width=\columnwidth]{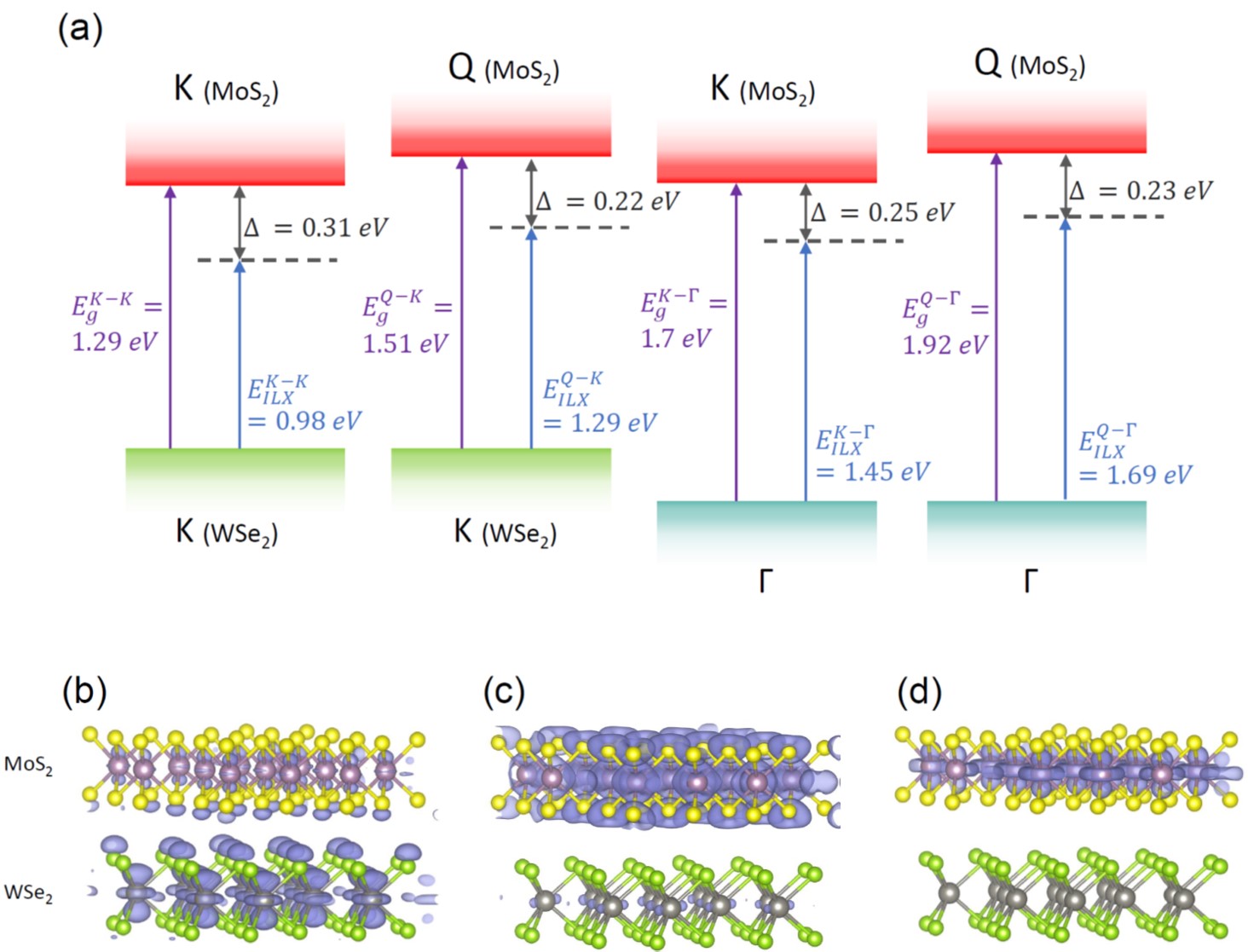} \caption{\label{fig:figureX} (a) Schematic illustration of relevant direct and indirect transitions between $\Gamma$, Q and K points from our \textit{ab initio} simulations. In case of K and Q, brackets indicate the sublayers the band edges belong to. $E_g$ are the \textit{electronic band gaps} from calculation with the HSE12 functional (see Supplementary Information S4). Dashed lines indicate energy positions of bound excitons with binding energies $\Delta$ obtained from solution of the excitonic Bethe-Salpeter Equation (BSE). (b) Plot of the {hole} wavefunction of the local VBM at the $\Gamma$ point of the heterostructure. (c) {Electron w}avefunction corresponding to the Q point local CBM of the MoS$_2$ sublayer. Notice a small spillover of the wavefunction into the WSe$_2$ layer. (d)  Plot of the {electron} wavefunction at the $K$ point {conduction band minimum (CBM) of the MoS$_2$} sublayer, which also is the global {CBM} of the heterostructure. 
} \end{figure}

To shed further light on the nature of ILE$_1$ and ILE$_2$, we performed additional theoretical calculations to estimate the binding energies of the momentum-indirect K-$\Gamma$, Q-$\Gamma$ and Q-K excitons combining the solution of the Bethe-Salpeter Equation {(BSE)} with density functional theory calculations. We refer to the method section and the Supplementary Information for details on the computational approach. 
Figure~\ref{fig:figureX}~(a) shows the obtained electronic band gaps, exciton binding energies and predicted excitonic peak positions of the direct K-K and the three candidate indirect transitions. 

For the interlayer K-K transitions, we predict a binding energy and a peak position of 310\,meV and 1\,eV, respectively, in good agreement with experiments~\cite{Karni_PRL2019} and previous theoretical works.~\cite{lantini-2017,Gillen-interlayer} According to our calculations, a Q-K exciton with binding energy 220\,meV should appear at an energy of about 1.3\,eV, however, to the best of our knowledge, it has not been observed experimentally so far. {K-$\Gamma$ and Q-$\Gamma$ excitons involve the same hole valley at $\Gamma$, but different electronic valleys with different corresponding effective masses. Therefore, the binding energies for the K-$\Gamma$ and Q-$\Gamma$ excitons may also be different. Nevertheless, we find very similar binding energies  for these excitons (Fig.~\ref{fig:figureX}~(a)), and the difference in PL peak positions $E_{ILX}$ is determined by the difference in electronic band gaps, such that $E_{ILX}^{Q-\Gamma}>E_{ILX}^{K-\Gamma}$.}
%For the excitons involving the $\Gamma$ point, we find very similar exciton binding energies and the difference in PL peak positions $E_{ILX}$ is determined by the difference in electronic band gaps, such that $E_{ILX}^{Q-\Gamma}>E_{ILX}^{K-\Gamma}$. 
This would suggest that ILE$_2$ indeed should be assigned to a K-$\Gamma$ exciton, while ILE$_1$ is the Q-$\Gamma$ transition. 
The energies of the ILE$_1$ and ILE$_2$ peaks (1.61 eV and 1.55 eV) are very close to the calculated values of $E_{ILX}^{Q-\Gamma}$ = 1.69 eV and $E_{ILX}^{K-\Gamma}$ = 1.45 eV, correspondingly. We note that our DFT calculations likely overestimate the energy difference between the K and Q conduction band minima compared to more sophisticated GW calculations, as observed in lattice-matched MoSe$_2$/WSe$_2$ heterostructures.~\cite{Gillen-interlayer}

\begin{figure} \centering \includegraphics[width=15cm,
    keepaspectratio]{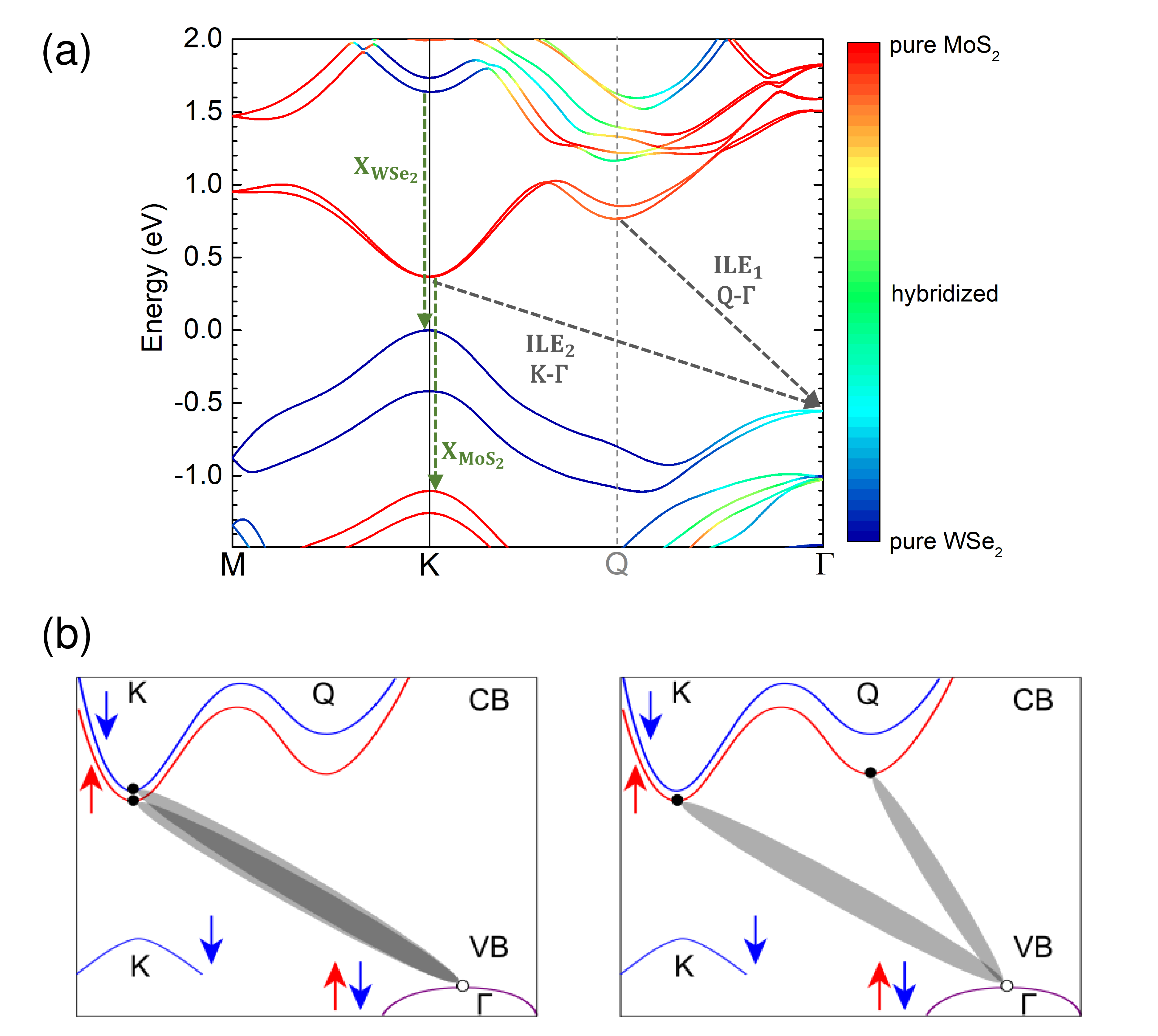} \caption{\label{fig:figure3} (a) DFT band structure of a MoS$_2$/WSe$_2$ heterostructure containing one formula unit of MoS$_2$ and WSe$_2$ used in our calculations (see Supplementary S4). The interlayer hybridization distribution is shown with the color code. The intralayer X$_{\rm{MoS2}}$ {and X$_{\rm{WSe2}}$}, as well as the momentum-indirect interlayer ILE$_1$ and ILE$_2$ transitions are illustrated. We note that the band energies are significantly affected by the strain, while the hybridization reasonably agrees with the results obtained for a strain-free heterostructure.(b) Schematic of the KK-$\Gamma$ and QK-$\Gamma$ trions forming upon electron doping.
     } \end{figure}

According to Fermi’s golden rule, the exciton PL intensity is proportional to the square of the transition matrix element, which, in turn, depends on the overlap between electron and hole wave functions. Momentum-indirect ILEs in type-II TMD heterostructures involve strongly hybridized valleys with large interlayer orbital overlap, and therefore, have significant PL intensity.~\cite{Kunstmann_Nature18, Okada2018, Wurstbauer_PRB20}  To compare the Q-$\Gamma$ and K-$\Gamma$ ILEs, we should have to look at the wavefunctions of electrons in Q and K valleys of MoS$_2$ and a hole in $\Gamma$ valley of WSe$_2$. Fig.~\ref{fig:figureX}~(b)-(d) demonstrates a certain delocalization of an electron in Q (panel (c)) and a hole in $\Gamma$ (panel (b)), while this delocalization is completely absent for electrons in the K points of the sublayers (panel (d) for the MoS$_2$ K point). The reason for this is that both Q and $\Gamma$ valleys in the MoS$_2$/WSe$_2$ heterostructure are hybridized, as shown by the color code in Fig.~\ref{fig:figure3} (a).
This hybridization is possible thanks to the energy overlap of MoS$_2$ and WSe$_2$ orbitals in the $\Gamma$ valley of VB and the Q valley of CB (see Supplementary Information, Fig. S5 (b)). Moreover, the corresponding MoS$_2$ and WSe$_2$ orbitals have similar character -- predominantly out-of-plane (\textit{d}$_{z^2}$) in VB $\Gamma$-valley and mixed in-plane (\textit{d}$_{x^2-y^2}$ and \textit{d}$_{xy}$) and out-of-plane in CB Q-valley .\cite{Liu_PRB2013, Kang_APL_2013, Shanavas_PRB_2015, Korm_nyos_2015}
The increased electron-hole wavefunction overlap for the Q-$\Gamma$ transition might explain the brightness of the ILE$_1$ peak in PL spectra compared to the weaker intensity of ILE$_2$ (corresponding electron-hole wavefunction overlap for the K-$\Gamma$ transition). 

\begin{spacing}{1.0}
%\singlespacing
\end{spacing}
%%%%%%%%%%%%%%%%%%%%%%%%%%%%%%%%%%%%%%%%%%%%%%%%%%%%%%%%%%%%%%%%%%%%%%%%%%%%%%%%

Let us discuss the difference between two momentum-indirect interlayer exciton Q-$\Gamma$ and K-$\Gamma$ with respect to the response to electron doping. Upon doping, donated electrons first populate the CB minimum, which is located in the K valley of MoS$_2$. The Q valley gets populated only at charge carrier density larger than 2 -- 6 $\times$ 10$^{13}$ cm$^{-2}$.~\cite{Piatti2018, Zhao_2020} As will be seen from our $\mu$-ARPES data shown below, we reach the electron densities when only the K valley gets populated. Additional charge carriers at the K valley lead to the conversion of K-$\Gamma$ interlayer excitons (ILE$_2$) into the trions with two electrons in the K (and/or K') valley and a hole in the $\Gamma$ valley. Let us denote them as KK-$\Gamma$ trions. As discussed above, the exciton-to-trion conversion and subsequent non-radiative recombination of trions is the most common mechanism of exciton suppression in TMDs upon charge carrier doping.  
Fig. 4 (b) illustrates the formation of KK-$\Gamma$ and QK-$\Gamma$ trions upon electron doping. For simplicity, we show only the KK-$\Gamma$ singlet trion, but there also should be a KK'-$\Gamma$ trion and the respective triplet variants with very similar energies, which might appear upon doping.
 A negative trion can be considered as two excitons sharing the same hole. That is, the QK-$\Gamma$ trion having two electrons in the Q and K valleys is the combination of Q-$\Gamma$ and K-$\Gamma$ excitons. Our calculations demonstrate that the K-$\Gamma$ ILE has a lower energy than the Q-$\Gamma$ ILE. Therefore, the KK-$\Gamma$ trion should also have a lower energy than the QK-$\Gamma$ trion. Since trions are fermionic states following the Pauli exclusion principle, the lowest energy ground state corresponding to the KK-$\Gamma$ trion will be populated first. As a result, the Q-$\Gamma$ excitons are more stable to the conversion into trions than the K-$\Gamma$ excitons, and, therefore, are more resistant to electron doping.  This confirms the experimentally observed stability of the ILE$_1$ peak in the Fig. 2.

\begin{spacing}{1.0}
%\singlespacing
\end{spacing}

\section*{The band alignment and $\mu$-ARPES studies}

An important question is the change of ILEs energy upon doping. Deposition of alkali metals on the surface of van der Waals layered materials usually leads to the strong out-of-plane potential difference (Stark effect), with the topmost layers being much more heavily doped than the bottom layers.~\cite{ Kang2017, Kim2017, Niels_PRB_Cs_BPh} If Cs atoms were resting on top of the MoS$_2$/WSe$_2$ heterostructure, the top MoS$_{2}$ monolayer would be much more doped than the bottom WSe$_{2}$ monolayer. This would lead to a strong modification of the band alignment, which, in turn, would result in a significant change of the ILEs energy. Our data in Fig. 2 show that the final doping level causes a shift of the ILE$_1$ peak by only 10 meV. The change in carrier concentration also affects the exciton binding energy and the quasiparticle band gap. Nevertheless, these two renormalization effects usually counteract each other, and the total change in optical band gap is minor.~\cite{Ugeda2014, Gao2016} Therefore, small changes in the ILEs energy suggest that the band alignment in the MoS$_2$/WSe$_2$ heterostructure should not be strongly affected by Cs doping. To understand this behaviour, we investigate our system by $\mu$-ARPES.

\begin{spacing}{1.0}
%\singlespacing
\begin{figure}\centering \includegraphics[width=16cm,
    keepaspectratio]{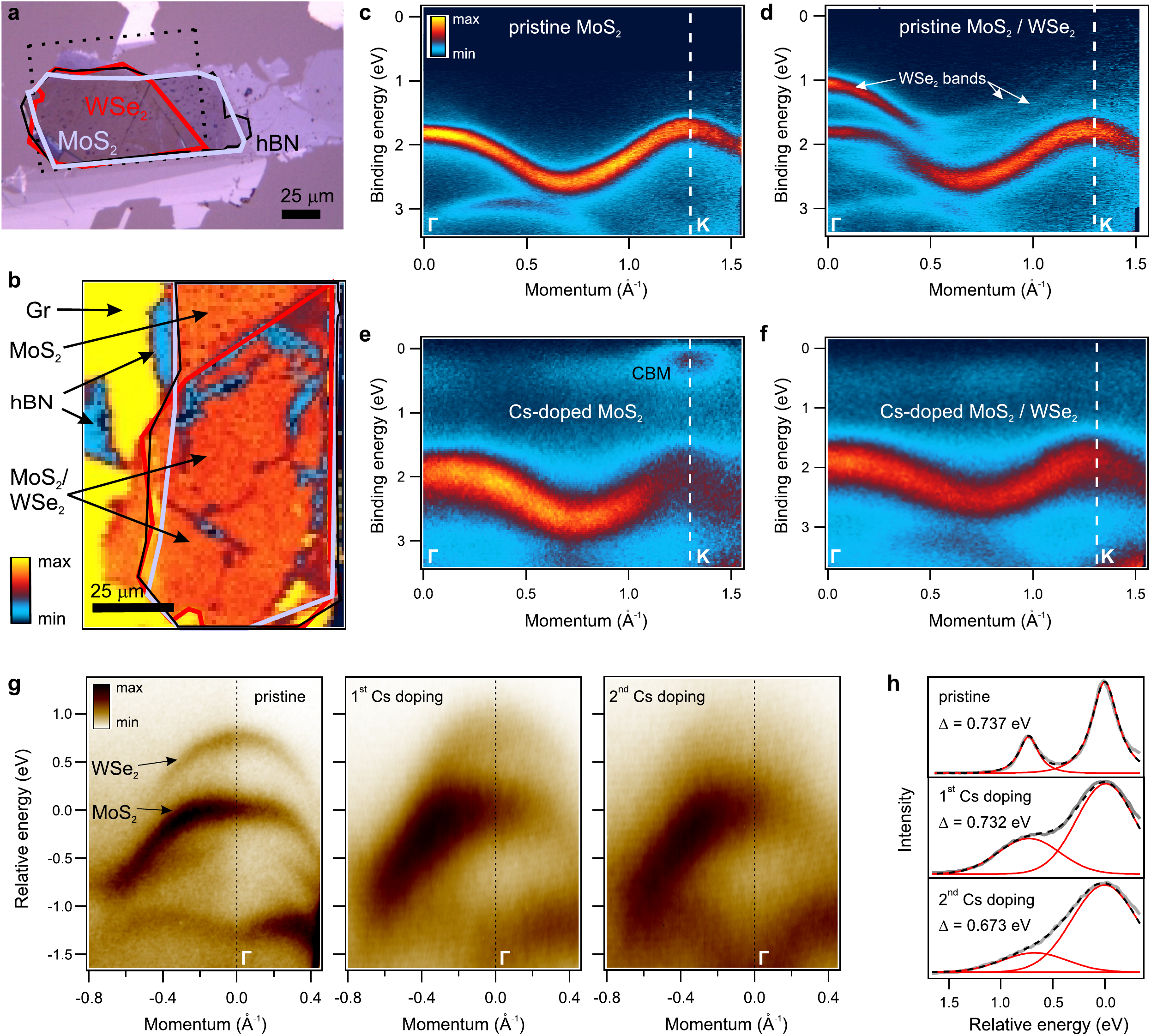} \caption{\label{fig:figure4} (a) Optical microscope image of the aligned MoS$_2$/WSe$_2$ heterostructure on hBN on epitaxial bilayer graphene on SiC and (b) its SPEM image, representing the integrated intensity of the $\mu$-ARPES image around the K point. MoS$_2$ is the upper layer, and WSe$_2$ is the lower layer. (c, d) Momentum slices of the $\mu$-ARPES data along the K-$\Gamma$ direction of the (c) monolayer MoS$_2$ and (d) MoS$_2$/WSe$_2$ regions. The energy bands attributed to the WSe$_2$ are denoted. (e, f) The same as in panels (c, d) but after Cs doping ($\sim$ 2 {\AA}). (g) $\mu$-ARPES spectra of the pristine and doped MoS$_2$/WSe$_2$ heterostructure on hBN/graphite/Si acquired along the K-$\Gamma$ direction. Each Cs dose was $\sim$ 1 {\AA}. The energy is taken relative the MoS$_2$ band at the $\Gamma$ point. (h) EDCs from the spectra in panel (g) at the $\Gamma$ point. The energy difference between the MoS$_2$ and WSe$_2$ bands is indicated as $\Delta$.} \end{figure}
\end{spacing}

In our experiments we compare MoS$_2$ monolayer and MoS$_2$/WSe$_2$ heterostructure doped by an identical amount of Cs. Fig.~\ref{fig:figure4} summarizes the $\mu$-ARPES data. The optical image of aligned MoS$_2$/WSe$_2$ heterostructure on an hBN flake on bilayer graphene/SiC used for the $\mu$-ARPES studies is shown in Fig.~\ref{fig:figure4} (a). The scanning photoemission microscopy (SPEM) image allows identification of the sample location, see Fig.~\ref{fig:figure4} (b). The SPEM image visualizes the integrated photoemission intensity around the K point of MoS$_2$ and WSe$_2$. Figs.~\ref{fig:figure4} (c) and (d) show the energy band structure of the monolayer MoS$_2$ and the MoS$_2$/WSe$_2$ heterostructure measured by $\mu$-ARPES on the corresponding sample regions. In the heterostructure region (Fig.~\ref{fig:figure4} (d)), in addition to the MoS$_2$ bands one can see bands of WSe$_2$ indicated by the arrows. Specifically, a WSe$_2$ band appears around the $\Gamma$ point and two spin-split bands appear with vanishing intensities as the K point is approached. In the heterostructure, the WSe$_2$ states at the $\Gamma$ point are more dispersing as compared to the states in monolayer WSe$_2$, where the topmost band at the $\Gamma$ point is almost flat (see Supplementary Information, Fig. S3). This is the result of interlayer hybridization with MoS$_2$ states, also predicted by our calculations (Fig. 3 (b)). Figs.~\ref{fig:figure4} (e) and (f) demonstrate corresponding spectra after deposition of $\sim$ 2 {\AA} of Cs adatoms on the sample. The monolayer MoS$_2$ becomes degenerately doped, as we observe the CB states corresponding to the K valley below the Fermi level, Fig.~\ref{fig:figure4} (e). This suggests a doping level of $\sim$ 10$^{13}$ electrons per cm$^2$.~\cite{Ehlen_2018} 
{Further deposition of Cs did not result in the appearance of a Q point below the Fermi level, supporting the fact that the PL signal from the Q-$\Gamma$ ILE was not completely suppressed (Fig. 2), and we are close the maximum possible Cs doping of our system.}
At the same time, the MoS$_2$/WSe$_2$ heterostructure is not turned into a metal, since its CB does not cross the Fermi level, Fig.~\ref{fig:figure4} (f). This implies that in the heterostructure the upper MoS$_2$ layer is less doped compared to the monolayer sample, and a significant amount of charge is transferred to the lower WSe$_2$ layer. We then performed an experiment with step-by-step deposition of Cs. The obtained high-resolution $\mu$-ARPES spectra are shown in Fig.~\ref{fig:figure4} (g), and the corresponding energy distribution curves (EDCs) at the $\Gamma$ point are shown in Fig.~\ref{fig:figure4} (h). After the first Cs dose ($\sim$ 1 {\AA}), the MoS$_2$ and WSe$_2$ bands have large energy broadening, but the energy distance between them (indicated as $\Delta$ in Fig.~\ref{fig:figure4} h) is not changed. The second deposition of the same Cs dose leads to the minor reduction of the $\Delta$. Therefore, we can conclude that the Cs doping has a minor impact on the relative energy positions of MoS$_2$ and WSe$_2$ bands in the heterostructure. This suggests an equal charge doping of MoS$_2$ and WSe$_2$ monolayers, which can occur when Cs atoms are intercalated into the interlayer van der Waals gap. This is also supported by our X-ray photoemission spectroscopy (XPS) data (see Supplementary Information, Fig. S6). A similar behaviour was previously observed by ARPES during the deposition of potassium onto the surface of a bulk MoS$_2$ crystal.~\cite{Eknapakul2014} Previous DFT calculations also supported that intercalation of deposited alkali metals into the van der Waals gap between the topmost TMD monolayers is much more energetically favorable than the adsorbtion on the surface.~\cite{Eknapakul2014} 

It should be noted that in addition to electron doping, which can be implemented in devices, other effects also take place during chemical doping. First, the presence of positive Cs ions can contribute to exciton scattering.~\cite{Auger_scattering_defects_PRB15}
Second, intercalation of alkali metals leads to an increased interlayer distance.
Nevertheless, our experimental data suggest that Cs intercalation does not completely suppress the Q-$\Gamma$ exciton (as occurs in the case of other excitons): its intensity is reduced by $\sim$ 3 times at largest Cs dose of $\sim$ 2 {\AA} (Fig. 2). Concerning the energy position of Q-$\Gamma$ exciton peak, theory predicts that the increase of interlayer distance from 0.6 nm to 1 nm should reduce the ILE binding energy in MoS$_2$/WSe$_2$ heterostructure by about 15 meV.~\cite{Donck_PRB_2018} This is a small change compared to the two competing factors -- the band gap renormalization and the reduction in exciton binding energy due to doping,~\cite{Ehlen_2018} and is comparable with changes that we observe in our data, were the final doping level causes a shift of the ILE$_1$ peak by only 10 meV.

\section*{Conclusions}
In conclusion, we fabricated high-quality MoS$_2$/WSe$_2$ heterostructures on hBN/graphene/SiC, and characterize their optoelectronic properties and the effect of chemical doping using combination of UHV PL and $\mu$-ARPES techniques. Our PL spectra revealed two interlayer excitons, {the energies of which agree very well with our \textit{ab initio} calculations of momentum-indirect K-$\Gamma$ and Q-$\Gamma$ optical transitions. The relatively high intensity of the Q-$\Gamma$ ILE is ascribed to the significan overlap of both electron and hole wave-functions of the hybridized Q and $\Gamma$ valleys.} To investigate the effect of chemical doping on the ILEs, we performed deposition of Cs atoms onto the MoS$_2$/WSe$_2$ system. Room-temperature UHV PL measurements reveal that electron doping induced by Cs leads to the suppression of the emission from both WSe$_2$ and MoS$_2$ intralayer excitons and the K-$\Gamma$ interlayer exciton. In contrast, PL from Q-$\Gamma$ interlayer exciton is robust to chemical doping, and is preserved even at high electron concentration ($\sim$ 10$^{13}$ cm$^{-2}$). 
{We attribute this to the stability of Q-$\Gamma$ ILE to the conversion into trions, which recombine non-radiatively at room temperature.} 
{Further insight into the interplay between radiative and non-radiative relaxation channels in chemically doped TMD heterostructures can be provided by time-resolved PL measurements.}
With $\mu$-ARPES we found that upon Cs doping the electronic band alignment is almost not changed. This suggests equal charge doping of both monolayers and intercalation of Cs in the interlayer gap. Two doping-induced effects, quasiparticle band gap renormalization and the reduction of exciton binding energy due to the increased screening, counteract each other. As a result, the energy shift of  Q-$\Gamma$ interlayer exciton upon doping is only 10 meV. The robustness of momentum-indirect interlayer excitons creates a new playground for optical studies of doped TMD heterostructures. This also opens an opportunity to use TMDs in light-emitting devices at a high concentration of charge carriers at room temperature. 

\section*{Methods}
\subsection{Fabrication of heterostructures}
{{MoS$_2$/WSe$_2$/hBN heterostructures were fabricated on epitaxial bilayer graphene on silicon carbide (SiC). Using bilayer graphene on SiC as a substrate allows us to perform both UHV PL and $\mu$-ARPES studies. We exfoliated MoS$_2$ and WSe$_2$ monolayers and hBN flakes onto viscoelastic PDMS stamps.~\cite{Castellanos-Gomez2014} The PDMS stamps prior to exfoliation were cleaned with a UV-lamp to minimize organic residues from polymers.~\cite{Novotny_PDMScleaning} The whole cleaning process took 15 minutes. The crystallographic axis (zigzag or armchair) of TMD monolayers on PDMS stamps was determined using the SHG measurements. Then we used a transfer system to create heterostructures layer by layer. First, we've transferred a thin (few tens of nm) hBN flake to the substrate by a PDMS clean stamp. Next, a WSe$_2$ monolayer was placed on hBN at a temperature of 60$^{\circ}$C. After that, we align the bottom WSe$_2$ layer on substrate with the PDMS stamp on which MoS$_2$ is located. After alignment, the MoS$_2$ was transferred on top of WSe$_2$. Finally, we annealed the samples in high vacuum at 250$^{\circ}$C for 3 hours.}} 

\subsection{PL and Raman measurements}

PL and Raman characterisation of MoS$_2$/WSe$_2$ samples was performed at room temperature in the back-scattering geometry using Renishaw inVia setup with 633 nm and 532 nm lasers. We adjust the laser power below 1 mW to not induce degradation of samples during the measurements. The laser spot size was about 1 $\mu$m. The low wavenumber Raman data were acquired using the Eclipse filter from Renishaw. UHV PL experiments during Cs doping were performed at room temperature using the same commercial setup with the laser light aligning in a home-built optical chamber.~\cite{Senkovskiy2017, Senkovskiy_NL2017} In our UHV PL setup we installed a motorized beam splitter with a mini camera and a light source. This allowed us to focus the laser onto the interesting sample area and accumulate PL spectra from the same sample spot after each Cs dose. Cs deposition was performed in the same chamber using a SAES getter. The Cs dose was controlled by a quartz crystal microbalance sensor. 

\subsection{$\mu$-ARPES experiments}
$\mu$-ARPES experiments were performed using synchrotron radiation facilities -- Spectromicroscopy beamline at ELETTRA (data presented in Fig. 5 c-f) and ANTARES beamline at SOLEIL (Fig. 5 g,h). In all experiments Cs was deposited on the sample at room temperature. The samples were aligned using the ARPES band dispersion maps to measure along the $\Gamma$-K direction. The data from the Spectromicroscopy beamline were measured with the photon energy of 27 eV at 40 K and a beam size of 0.6 $\mu$m. The data from the ANTARES beamline were measured with the photon energy of 100 eV at 80 K and a beam size of 0.7 $\mu$m. Before the measurements the samples were annealed at 250$^{\circ}$C -- 300$^{\circ}$C for 3 -- 6 hours. The Cs deposition was also controlled by a quartz crystal microbalance sensor.

\subsection{Computational methods}

To predict exciton peak energies despite the large lattice mismatch, we followed the two-pronged approach previously reported in Ref.~\cite{Karni_PRL2019}: To obtain the relevant electronic band energies and band gaps, we built an nearly strain-free supercell (<0.05\% strain) with a 16.1$^\circ$ twist angle between the MoS$_2$ and the WSe$_2$ layers. We used the QUANTUM Espresso suite~\cite{qe} and the PBE+D3 exchange-correlation (XC) approximation to optimize the atomic positions and interlayer distances, while keeping the lattice constant fixed at the average of the lattice constants of the isolated systems. The HSE12 XC functional~\cite{hse12} was then used to derive the electronic band energies of interest. Spin-orbit interaction was fully included in the calculation of the electronic structure. 
The exciton binding energies were calculated using a minimal unit cell containing one formula unit of both MoS$_2$ and WSe$_2$, where the individual layers were strained by about 2$\%$. %While the strain strongly affects the electronic band gaps, we found the exciton binding energies much less affected. 
Based on the electronic structure obtained from DFT calculations, we used a modified version of the YAMBO code~\cite{yambo,gillen-pssb-2021} to calculate the binding energies of direct and indirect excitons of interest from solution of the excitonic Bethe-Salpeter Equation. The excitonic transition energies were then estimated from a combination of the electronic band gaps from HSE12 calculations and the exciton binding energies. We refer to section 4 of the Supplementary Information for details on the computational parameters used for the DFT and BSE calculations.

\section*{Data availability}
The datasets generated during and/or analysed during the current study are available from the corresponding authors on reasonable request.

\section*{Supplementary information}
Supplementary Information: Additional experimental details, including sample preparation and characterization methods; Computation methods and details.

%\section*{Author contributions}

\begin{addendum}
\item[Competing Interests] The authors declare no competing interests.
   
\item[Funding Sources]
B.V.S. acknowledges DFG project SE2575/4-1 'Engineering the electronic band structure of transition metal dichalcogenide heterostructures in device geometries'.
A.D., O.N.G.L., B.V.S. and A.G. acknowledge support through the CRC 1238 within project A01 and the ERC grant no. 648589 'SUPER-2D'.

\item[Correspondence] Correspondence and requests for materials should be addressed to EK and BS (email: e.khestanova@gmail.com, senkovskiy@ph2.uni-koeln.de).
\end{addendum}

%%%%%%%%%%%%%%%%%%%%
%%% BIBLIOGRAPHY %%%
%%%%%%%%%%%%%%%%%%%%

\bibliographystyle{naturemag}
\bibliography{bibliography_TMD}
\clearpage
\begin{figure}\centering \includegraphics[width=18cm,
    keepaspectratio]{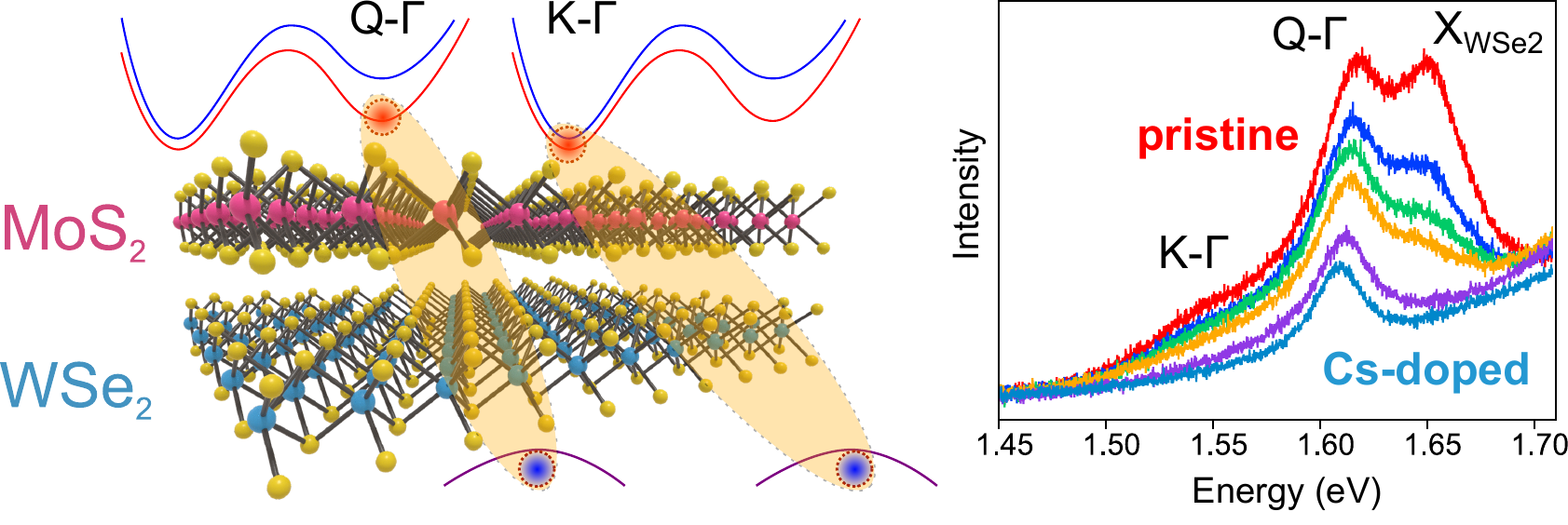} \caption{TOC} \end{figure}
\end{document}